\documentclass{kluwer}    

\usepackage{epsfig}

\newdisplay{guess}{Conjecture}

\newcommand{\citeP}[1]{(\citeauthor{#1} \cite*{#1})}
\newcommand{\citeN}[1]{\citeauthor{#1} (\cite*{#1})}
\newcommand{\citeNP}[1]{\citeauthor{#1} \cite*{#1}}

\newcommand{\apj} {ApJ}
\newcommand{\apjs} {ApJS}
\newcommand{\apjl} {ApJL}
\newcommand{\aap} {A\&A}
\newcommand{\solphys} {Solar Physics}
\newcommand{\arcsec} {''}

\begin{document}

\begin{article}
\begin{opening}         
\title{Motivation and Initial Results from SPINOR in the Near Infrared}
\author{Hector \surname{Socas-Navarro}, David \surname{Elmore} \&
  Bruce W. \surname{Lites}   \thanks{Visiting Astronomers, National Solar
    Observatory, 
    operated by the Association of Universities for Research in Astronomy,
    Inc. (AURA), under cooperative agreement with the National Science
    Foundation.} }
\runningauthor{Socas-Navarro, Elmore \& Lites}
\runningtitle{Results from SPINOR in the Near Infrared}
\institute{High Altitude Observatory, NCAR \thanks{The National Center for
  Atmospheric Research (NCAR) is sponsored by the National Science
  Foundation}} 
\date{October 1, 2004}

\begin{abstract}
SPINOR is a new spectro-polarimeter that will serve as a facility 
instrument for the Dunn Solar Telescope at the National Solar
Observatory. This instrument is capable of
achromatic operation over a very broad range of wavelengths, from $\sim$400
up to 1600~nm, allowing for the simultaneous observation of several
visible and infrared spectral regions with full Stokes polarimetry. Another
key feature of the design is its flexibility to observe virtually any
combination of spectral lines, limited only by practical considerations
(e.g., the number of detectors available, space on the optical bench,
etc). SPINOR is scheduled for commissioning by the end of 2005. In this
paper we report on the current status of the project and present actual
observations of active regions in the Ca~II infrared triplet and the He~I
multiplet at 1083~nm.
\end{abstract}
\keywords{instrumentation: polarimeters, polarization, telescopes, Sun:
  magnetic fields,   Sun: photosphere, Sun: chromosphere}

\end{opening}           

\section{Introduction}
\label{sec:intro}

The new breed of
spectro-polarimeters scheduled to begin operation in the next
few years make the present a very exciting time for solar physics. 
Examples of these instruments are SOLIS \citeP{K98}, Solar-B \citeP{LES01},
POLIS \citeP{SBK+03},
Sunrise \citeP{SCG+03}, and the DLSP \citeP{SEL+03}. All these instruments,
however, are highly specialized and operate under fixed conditions, either
with the aim of performing synoptic observations or to optimize for spatial
resolution. While these aspects are very important for the future advance
of solar physics, we feel that there is also a need for an
``experiment-oriented'' type of instrument. By this we mean an instrument
that provides enough flexibility to implement more or less arbitrary optical
arrangements at the observer's request. Such an instrument should allow
researches to observe any given spectral line (or combinations of lines),
either at the disk or off the limb at a diverse range of wavelengths. One
should be able to modify parameters like spatial resolution, spectral
dispersion, integration times, etc, in order to address a broad range of scientific
problems. In some sense, operating an instrument of this kind is
like setting up an experiment in a laboratory. Optical components and 
detector systems are arranged on an optical bench to investigate some
specific problem. 

For more than a decade, the highly successful Advanced Stokes Polarimeter
(ASP, \citeNP{ELT+92}) has provided these capabilities at {\it visible}
wavelengths. The 
Spectro-Polarimeter for Infrared and Optical Regions (SPINOR) will
replace the ASP as the experiment-oriented spectro-polarimeter on the
Dunn Solar Telescope (DST) at the National Solar Observatory (NSO), Sunspot,
NM, USA. SPINOR is currently under deveopment by the High Altitude
Observatory and the National Solar Observatory. 
It has been conceived with flexibility as
a top priority and intends to fill this gap in the next-generation solar
instrumentation. The new instrument will replace and enhance the capabilities
of the ASP.

Table~\ref{table1} below provides some specifications for SPINOR. The most
important enhancement over the ASP is the ability to observe simultaneously
in the visible {\it and} the near infrared. Higher spatial resolution than was
available over most of the life of the ASP is now
possible thanks the the high-order adaptive optics (AO) system \citeP{R00}
developed for the DST. Two different detectors will
be available at commissioning time (scheduled for the end of 2005): a
PixelVision Pluto 652x488 thinned back illuminated split frame
transfer CCD, 60 frames per second readout rate (hereafter Pluto) and a
Sarnoff CAM1M100, 1024x512, thinned back illuminated split frame
transfer CCD, 100 frames per second readout rate.
These will be fully dedicated for 
observations between 400 and $\sim$900~nm. The SPINOR optics allows
for an even broader range of wavelengths, up to 1.6~$\mu$m. We plan to take
advantage of this in future observations by using
an HAO Rockwell infrared camera that is shared with other HAO
projects. The modular design of the instrument control will allow for
additional detectors to be attached to SPINOR using NSO's Virtual
Camera interface.

The present configuration of the instrument, used in the engineering run
reported here, does not include the Sarnoff or the Rockwell cameras. For
this run we used the ASP control computers and detectors 
alongside the Pluto camera. With this configuration we were able
to demonstrate SPINOR's achromatic spectro-polarimetry up to 1083~nm. 

This paper is organized as follows. Section~\ref{sec:science} discusses the
scientific motivations for the development of
SPINOR. Section~\ref{sec:instrument} presents a description of the
polarimeter with some details on the new achromatic optics. The polarimetric
calibration of the system consisting of the 
telescope and the polarimeter is explained in section~\ref{sec:calib}. Some
observations taken during an engineering run in June 2004 are presented in
section~\ref{sec:obs}. Finally, section~\ref{sec:conc} presents the main
conclusions of our work and some future perspectives.

\section{Scientific Rationale}
\label{sec:science}

The main scientific driver behind the development of SPINOR is to expand
the present ASP magnetometric capabilities into the higher solar atmosphere
(chromosphere and corona). However, its infrared coverage is also
of great relevance for the study of problems associated with photospheric
magnetism. Below we summarize some of the major scientific problems for which
SPINOR is particularly suited, which cannot be pursued using the ASP (or
other existing instruments). However, being that this an experiment-oriented
instrument designed for maximum versatility, its most important applications
might be some that we cannot envision at this time 

\subsection{Photospheric magnetism}

Recent investigations of photospheric fields in the infrared often
reveal a different picture from conventional visible observations. A
remarkable example 
is the finding of supersonic flows in the penumbrae of sunspots by
\citeN{dTIBRC01} in data from the Tenerife Infrared Polarimeter (TIP). Such
strong flows had never been observed in 
the visible, with the exception of the peculiar $\delta$-configuration
sunspots (\citeNP{MPLS+94}), and flows near pores (\citeNP{LES01}).

Perhaps the most puzzling observations of photospheric fields in the infrared
are those of the quiet sun, particularly outside the magnetic network (the
region sometimes referred to as the photospheric internetwork).
Visible
observations indicate that most of the fields in this region are strong
($\sim$1.3~kG), but concentrated into
very small areas ($\sim$1\% of the pixel; see \citeNP{SNSA02} and references
therein). However, 
recent infrared observations (\citeNP{L95}; \citeNP{LR99}; \citeNP{KCS+03})
suggest that most of the fields are weak ($\sim$400~G) and diffuse. 
These contradictory results have 
sparked a controversy on the true nature of quiet sun fields. This issue
is an important one, since our current understanding indicates that most of the
solar magnetic flux (even at solar maximum) is located in the quiet sun
outside of active regions, and it is likely that the evolution of this flux
plays a role in the heating of the upper atmosphere.
A possible solution to the observational contradiction has been proposed by
\citeN{SN03b} and 
\citeN{SNSA03}. It turns out that the observations may be explained easily by a
small-scale distribution of fields, beyond 
the spatial resolution of the observations, having intermixed
weak and strong fields.
Moreover, these authors showed that the actual sub-pixel
distribution of the field can be inferred from simultaneous visible and
infrared observations
(like those from SPINOR).

The results of \citeN{SNSA03} may be extrapolated to other
physical scenarios in which different field strengths coexist within the
resolution element of the observations. Another example where this
strategy would be very useful is the investigation of sunspot penumbrae. It
is presently believed (see, e.g., \citeNP{S02}; \citeNP{SAB98}) that a
penumbra is formed by a large number of thin 
radial filaments embedded in a magnetic environment, with the filaments
having weaker and more horizontal fields and channeling the cool
siphon flows traditionally observed as the Evershed effect. The actual size,
properties and origin of these filaments are still a subject of intense
debate (\citeNP{MP00}; \citeNP{SA01b}; \citeNP{MP01}), but it seems well
established that they are not spatially 
resolved in the observations. The ability to infer spatially-unresolved
distributions of the magnetic field from simultaneous visible and infrared
observations would provide important clues on the structure of the penumbra.

Innovative new diagnostics of solar magnetic fields are emerging as a result
of parallel theoretical and observational advances.
In a recent effort to understand the anomalous polarization signals observed in
some spectral lines, \citeN{LATC02} studied the hyperfine
structure induced in the atomic energy levels by the nuclear spin, and its
effects on the polarization transfer process. They demonstrated that the
signature imprinted by the hyperfine structure on some spectral lines has an
important potential for magnetic field diagnostics. Some of the most
interesting lines lay in 
the wavelength domain between 800 and 1500~nm. Examples are the
Rb~I~D1 line at 794~nm , which shows a combination of hyperfine
structure and isotopic mix, or the Mn~I lines at 870~nm, and in the
infrared at 1.29, 1.33 and 1.52~$\mu$m.

\subsection{Chromospheric and coronal magnetism}

A new global picture of solar magnetism is emerging from
the seemingly disparate observational domains of photospheric small-scale 
magnetic fields and the diffuse, voluminous magnetic structure of the solar
corona.  The observation and interpretation
techniques used for photospheric and coronal studies are markedly
different. To further the development of this global view, we identify
a key missing ingredient: an in-depth investigation of the
interface layer, the chromosphere. Observational capability for chromospheric
vector magnetic fields and the associated dynamics has been lacking because most
interesting and/or useful lines
lie outside the wavelength coverage range of most solar
polarimeters (it should be noted, however, that
successful investigations of prominence fields have been carried out
recently; e.g.  \citeNP{CLA03}). Observable lines either form 
too low in the chromosphere (e.g., the Mg~I $b$-lines) or their
polarization transfer is still not well understood (e.g., H$_{\alpha}$). 

SPINOR would open new perspectives for chromospheric investigations with its
ability to observe the full polarization state of the 
Ca~II infrared triplet, around 854~nm. These
lines are the best candidates for chromospheric diagnostics, at least in the
Zeeman regime, due to their relatively simple formation physics, their long
wavelengths (which results in stronger Zeeman signals), and the
valuable information they carry on the thermal and magnetic
conditions of the higher atmosphere (\citeNP{U89}; \citeNP{SNTBRC00a};
\citeNP{SNTBRC00b}). They are also
sensitive to the Hanle effect, which 
provides complementary diagnostics on the weaker ($\sim$1~G) fields, and have
been successfully modeled by \citeN{MSTB01}.

Finally, the He~I multiplet at 1083~nm is of great interest for
chromospheric and coronal studies. This line is seen in emission in
prominences and in absorption in filaments, with strong polarization signals
arising from both Hanle and Zeeman effect. SPINOR would be able to
provide full spectro-polarimetry at 1083~nm,
which implies the potential to investigate the magnetic and
dynamic conditions of these structures. Other interesting coronal lines that
may be accessible for observations are the two Fe~XIII lines at 1074~nm,
although these may not be visible using a traditional solar telescope.

\section{Description of the instrument}
\label{sec:instrument}

SPINOR is based on the design of the ASP, which uses a rotating waveplate as
a modulator and a polarization beam splitter as a dual-beam analyzer. All of
the polarization optical components, however, have been replaced by new
achromatic ones. SPINOR utilizes the ASP calibration/modulation unit
(cal/mod) at the exit port of the DST, high-order AO, and the Horizontal
Spectrograph. 

Since SPINOR operates over a much wider wavelength range compared to ASP,
it was deemed necessary to allow SPINOR to use a selection of gratings.
Depending upon the desired spectral line
and spectral resolution, different gratings may be
selected. Higher blaze angle gives higher spectral resolution for the same
spatial sample size. Since SPINOR will operate with AO, its design favors
a higher spatial resolution than the ASP.  Since it is a general-purpose
research instrument, it cannot be optimized for a single line, so
the SPINOR spatial resolution is not as high as that attainable by the new DLSP.
These spectrograph issues lead typically to a spectrograph using a 40~$\mu$m
slit and 1000~mm camera lens. The observations reported in this paper were
carried out using the 308.57~lines/mm grating with a blaze angle of 52$^{\rm
  o}$.

\subsection{Specifications}

SPINOR has been conceived keeping versatility as the highest priority in
order to allow for a broad range of potential applications. The project is
being developed in a way that allows the ASP to remain operational until the
new instrument is completed. Its most important features are:

\begin{table*}
\caption[]{Performance comparison between SPINOR and ASP}
\label{table1}
\begin{tabular}{lccc}
\hline
Parameter &  & ASP & SPINOR \\
\hline
Calibratable wavelength range (nm) &  & 450-750 & 400-1600 \\
Field of view along slit (arc seconds) &  & 80 & 120 \\
Quantum efficiency & 400 nm & 0.01 & 0.72 \\
                   & 700 nm & 0.32 & 0.80 \\
                   & 1080 nm & 0.00 & 0.03, 0.60$^{*}$ \\
                   & 1600 nm & 0.00 & 0.60$^{*}$\\
                   &            &      & $^{*}$(with suitable IR camera)\\
Read noise (electrons) & & 50 & 25 \\
\hline
\end{tabular}
\end{table*}

\begin{itemize}
\item Achromatic optics from 400 to 1600~nm offer the
  capability of simultaneous observations at diverse wavelengths
  This extended range represents an important
  improvement over that of the ASP. Its potential scientific
  benefits have been discussed in more detail in section~\ref{sec:science}.
\item Detectors with higher quantum efficiency and lower noise than those
  in ASP allowing for higher signal to noise observations.
\item Use of state-of-the-art components, electronic systems, computers,
  detectors and software. Deployed over 12 years ago, ASP's technology is now
  15 or more years old, and some of its systems are starting to fail. Both routine 
  maintenance tasks and normal operation are compromised because many
  replacement parts are no longer available. The SPINOR system 
  will be more stable and suffer significantly less
  downtime. Data products will also be easier to handle 
  (e.g., using DVDs instead of magnetic tapes, simpler
  analysis procedures, etc), an improvement that will greatly facilitate
  the science and make the instrument more accessible to a broader user
  community.
\item Open and modular optical design, with room on the optical bench to incorporate
  and/or 
  replace components. SPINOR will be deployed at the DST as
  a set of instrument modules with ``virtual'' cameras, a concept that NSO
  has developed for the DLSP and other DST instrumentation. 
  This will allow diverse and complex observations,
  combining SPINOR with other DST instruments. The control software will be
  fully 
  customizable for a broad variety
  of observing modes (although several pre-defined modes will exist for
  frequently used configurations).
\end{itemize}

\subsection{Calibration}
\label{sec:calib}

The calibration procedure for SPINOR consists in the determination of the
Mueller matrices of the telescope (${\bf T}$) and the polarimeter (${\bf
  X}$). Any solar Stokes vector ${\bf s_{in}}$ is observed as ${\bf
  s_{out}}$: 
\begin{equation}
{\bf s_{out}} = {\bf XT s_{in}} \, .
\end{equation}

The Mueller matrix of the polarimeter depends on the particular instrumental
configuration and varies significantly in between observing runs (sometimes
even from one day to the next). For this reason, polarimeter calibration
operations are done typically once a day, or even more often if changes are
made to the instrument. Fortunately, calibration operations do not take too long
($\sim$20 minutes). A calibration linear polarizer and retarder are mounted
following the telescope exit window. These are used to polarize the light
beam in various known states. The known input vectors and the observed
outputs are used in a least-squares fitting procedure 
to determine the matrix elements of ${\bf X}$. This is essentially the same
procedure that was used for the ASP calibration, which is explained 
in the paper by \citeN{SLMP+97} and will not be repeated here. The only
difference is that the calibration
polarizer and retarder are now almost entirely achromatic over the broad
wavelength range of SPINOR. The retardance of the calibration retarder and
the orientation of its fast axis are taken as 
free parameters of the fit, in order to account for wavelength
variations and possible mounting inaccuracies. 

The telescope calibration requires an entire day of observations at different
positions of the Sun on the sky. This needs to be done at least every time
the turret mirrors are recoated, and desirably every few months.
We have devised a new method that departs significantly from that of
\citeN{SLMP+97}. The
telescope is modeled as a function of 9
free parameters: the entrance window retardance and axis orientation; the
turret mirrors retardance and ratio of reflectivies ($rs/rp$); the main
mirror retardance and $rs/rp$; the exit window retardance and axis
orientation; and finally the relative rotation between the telescope and the
polarimeter (see \citeNP{SLMP+97} for an explanation of these
parameters). 

An array of achromatic linear polarizers, placed in front of 
the entrance window, is used to introduce known states of polarization in the
system. These vectors are measured at various times of the day, covering a
wide range 
of the three variable angles of the telescope: turret azimuth, elevation
and table rotation. Again, a least-squares fit is used to determine the free
parameters of the model.

\begin{figure*}
\includegraphics[width=1.2\maxfloatwidth]{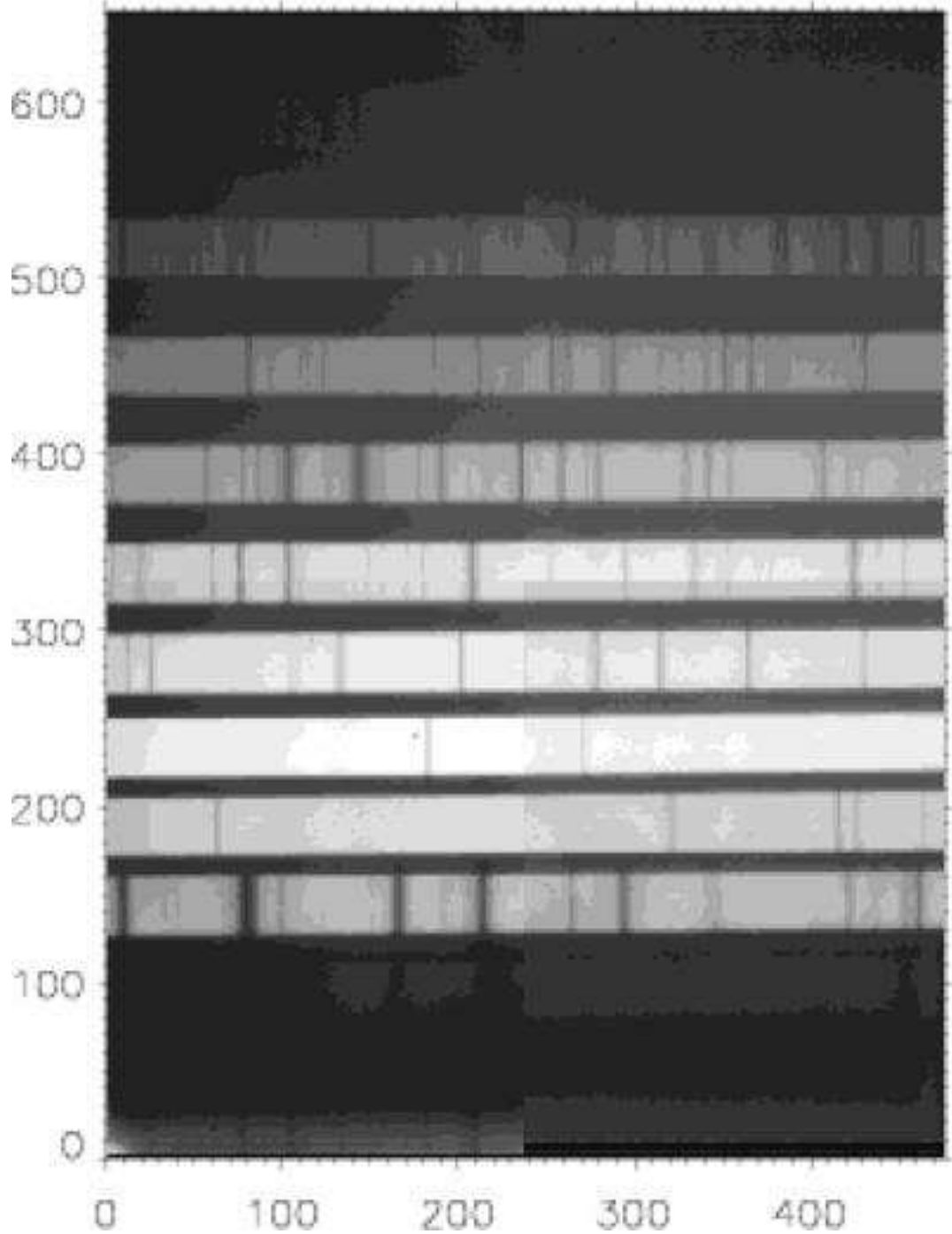}
\caption{
Cross-dispersed spectra from a telescope calibration operation. Central
wavelengths in nm from top to bottom: 416, 451, 492, 541, 601, 677, 773,
902. (Note: Figure has been downsampled for Astro-ph)}
\label{fig:tels}
\end{figure*}

The new SPINOR technique uses a cross-dispersing prism placed before one
of the detectors to record the  
various overlapping orders simultaneously (shown in Fig~\ref{fig:tels}). In
this manner we are able to 
obtain telescope calibration data for many wavelengths at once. Instead of
fitting each set of parameters independently for each wavelength, we take a
fully-consistent approach whereby the entire dataset is fit using a
wavelength-dependent model. Some parameters (window
retardances and mirror properties) are allowed to vary with
wavelength in the fit, up to the third order in $\lambda$. The rest of the
parameters 
are wavelength-independent. This model has sufficient freedom to reproduce
the measured vectors in the entire dataset and at the same time provides a
consistent model of the telescope. Table~\ref{table:tel} shows the various
telescope parameters obtained from our calibration at three different
wavelengths.

\begin{table} %
\begin{tabular}{lccc}                                        
\hline
Wavelength (nm)  &  400   &   600   &   800  \\
\hline
EW retardance  &  2.1   &  -4.3   &  -1.2   \\
EW orientation &   12.5   & 12.5    &  12.5     \\
Turr. rs/rp    &  1.12  & 1.08    & 1.11  \\
Turr. retardance & 145.1 & 160.1  & 163.4  \\
Main rs/rp     &  1.00  & 1.00    & 1.00   \\
Main retardance     &  179.9  & 179.9    & 179.9   \\
XW retardance  &  -9.05  &   -3.7   &   -3.1   \\
XW orientation  &  20.9   &  20.9   &   20.9  \\
T-X rotation   &  93.3   &  93.3   &  93.3  \\
\hline
\end{tabular}
\caption[]{DST telescope parameters retrieved with our calibration procedure.
EW: Entrance window, Turr: turret mirrors, XW: Exit window. All angles are
degrees. 
}\label{table:tel}
\end{table}

\section{Observations}
\label{sec:obs}

The observations presented in this paper were obtained during an engineering
run on June 2004 with the following setup. The ASP
cameras were configured to record the Ca~II lines at 849.8 and 854.2~nm
(plus some other photospheric and telluric lines blended in the wings of the
Ca lines). The new Pluto camera was observing at 1083.0~nm, including the
He~I multiplet, two photospheric lines and a telluric line. We aligned the
camera chips so that the broad Ca lines would fall near the edge of the
chip. In 
this manner we can record some wing ``continuum'' on the opposite side of the
detector, which is useful for the data analysis (polarization
calibration, flat-fielding and also as a reference for the thermodynamics in
the inversions). Since most of the 
polarization signal is concentrated near the line core, it is not necessary to
record both wings of the Ca lines. An additional advantage is that we can
also record a strong photospheric line in the 854.2~nm images.

The old ASP modulator was used because at the time of the observations we had
not yet received the new achromatic modulator from the manufacturer. This
resulted 
in less than optimum polarimetric efficiency at these long wavelengths,
especially at 1083~nm. The new modulator arrived later in the observing run
and 
is now available for use at the DST. 

\subsection{Magnetograms}

\begin{figure*}
\includegraphics[width=1.2\maxfloatwidth]{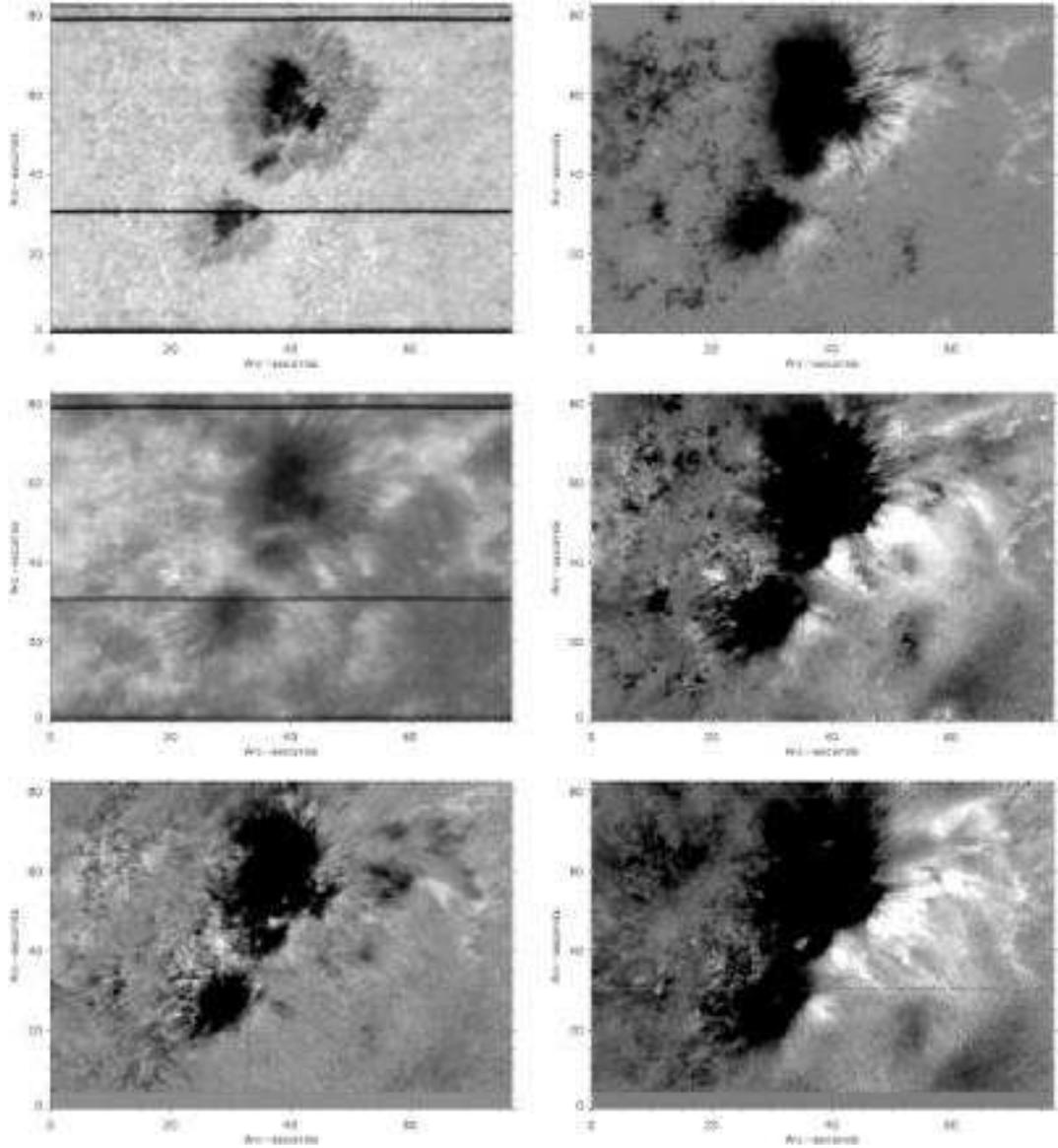}
\caption{
Various maps of active region NOAA~0634. From top to bottom, left to right:
Continuum intensity. Photospheric magnetogram. Chromospheric filtergram in
the core of the Ca~II 854.2~nm line. Chromospheric magnetogram taken on
the red side of the 849.8~nm line core. Chromospheric magnetogram on the
blue side of 854.2~nm. Chromospheric magnetogram on the red side of 854.2~nm
. The gray scale in all chromospheric [photospheric] magnetograms is $\pm$2\% 
[$\pm$3\%] of the corresponding Stokes I intensity (see text). Disk center is
towards the left of the image. (Note: Figure has been downsampled for Astro-ph)
}
\label{fig:maps}
\end{figure*}

\begin{figure*}
\includegraphics[width=1.2\maxfloatwidth]{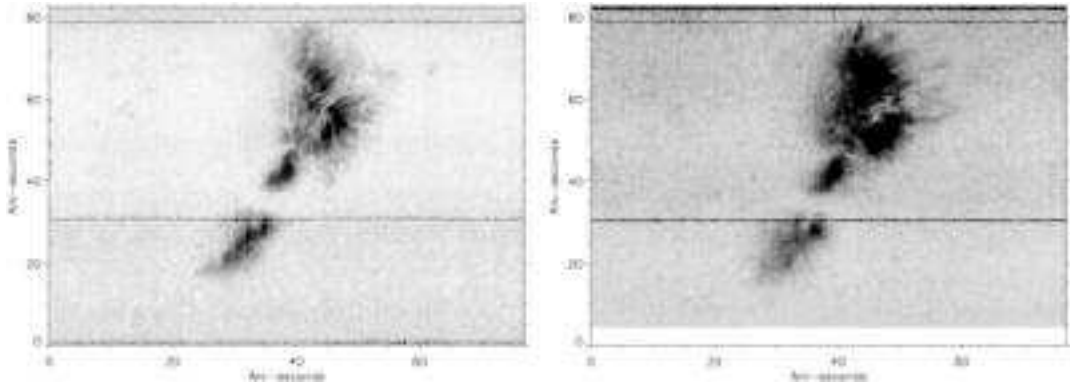}
\caption{
Magnetograms of linear polarization ($\sqrt{Q^2+U^2}$). Left:
Photospheric line (saturated at $\pm$4\%). Right: Chromospheric Ca~II line
at 854.2~nm (saturated at $\pm$2\%). (Note: Figure has been downsampled for Astro-ph)
}
\label{fig:linmaps}
\end{figure*}

Figs~\ref{fig:maps} and~\ref{fig:linmaps} show several maps of active region
NOAA~0634 observed on June 16 at 15:16~UT. This dataset has the best seeing
among our active region observations, with a 
granulation contrast of 3.3\% (notice that this value may not be directly
comparable to that of visible observations because of the variation of the
source function with wavelength). Some features apparent in the
magnetograms 
or the chromospheric filtegrams (e.g., the transversal fluctuations in the
penumbral filaments near coordinates [$x=50$,$y=55$] in the figure) exhibit
spatial scales as small as 0.7\arcsec. 

The maps represented in the figures have been obtained from a
spectro-polarimetric scan of the region consisting of 350 steps of
0.22\arcsec each. The 
scanning step oversamples the resolution element. This was done so that it
would be possible to bin the 10830 data in order to build up its photon count
without saturating the other two detectors. Unfortunately the 1083~nm
data from this day was unusable due to problems with the camera. Therefore we
only show the 849.8 and 854.2~nm regions for this dataset.  

\begin{figure*}
\includegraphics[width=1.2\maxfloatwidth]{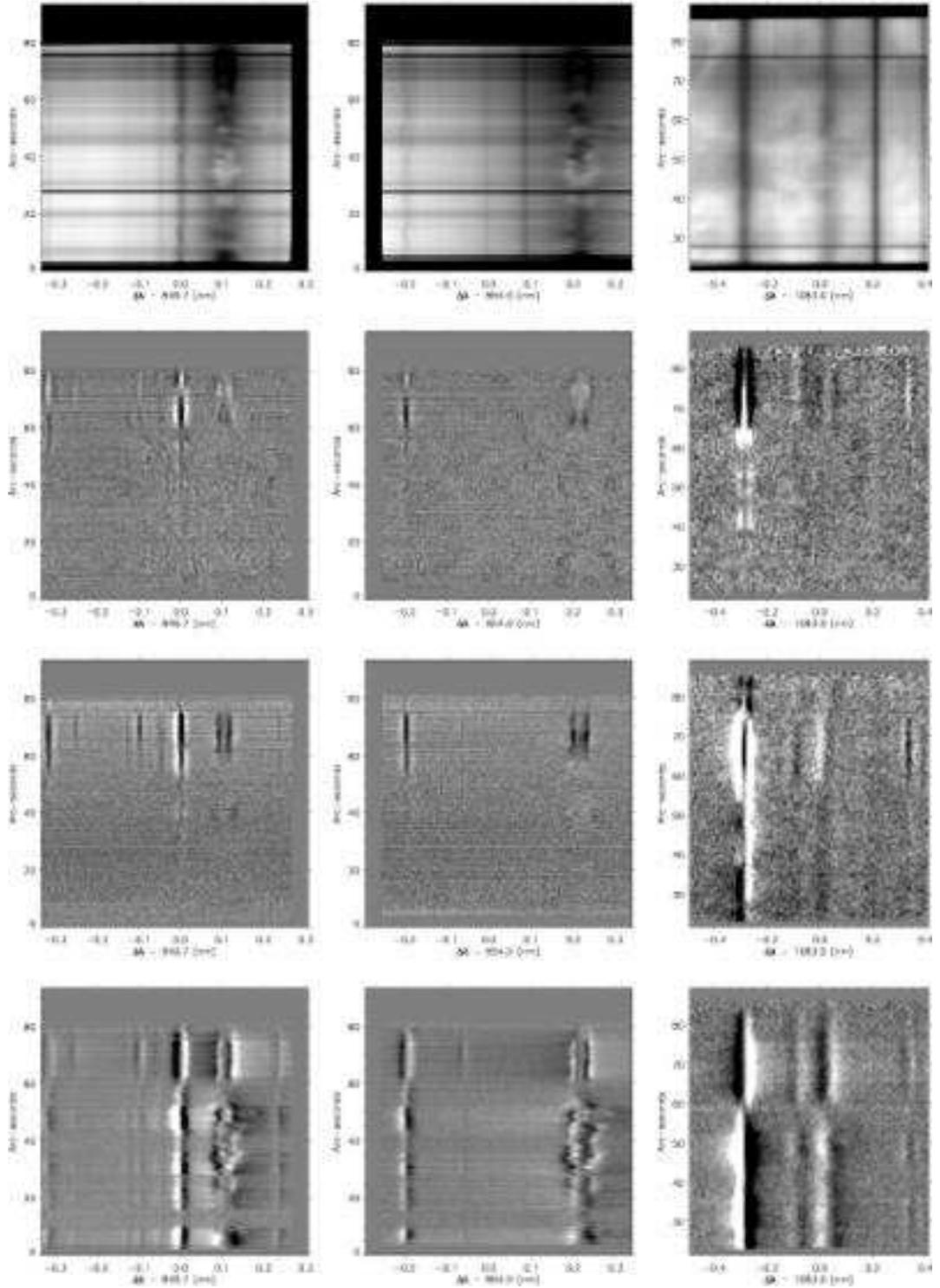}
\caption{
Stokes profiles from one particular slit position of the NOAA~0635
scan. Rows: Stokes I, Q, U and V parameters (from top to bottom). Columns:
Images from the camaras at 849.8 (left), 854.2 (middle) and 1083 (right) nm.
Profiles are normalized to quiet Sun continuum. Stokes Q and U panels are
saturated at $\pm$1\%, Stokes V panels are saturated at $\pm$2\%.(Note: Figure has been downsampled for Astro-ph)
}
\label{fig:profs}
\end{figure*}

The term ``magnetogram'' is used in this work somewhat inappropriately (but
following a very widespread terminology) to refer to the amount of
Stokes V signal integrated over a certain bandwidth on either the red or blue
side of the line center, and normalized to the intensity integrated over the
same bandwidth. We have used a 10-pixel wide square filter
function. Notice that this quantity is not necessarily related to the
magnetic field strength in the Sun. In some cases it may bear some
resemblance to the longitudinal magnetic flux density (as defined by
\citeNP{GLASN+02}), but this is not always the case (\citeNP{SN02})
especially given the complex polarization patterns exhibitted by the
chromospheric lines (see Fig~\ref{fig:profs}). For this reason, the red
and blue magnetograms represented in Fig~\ref{fig:maps} are different. A
reliable determination of the magnetic properties of the atmosphere would
require the application of a realistic inversion technique, such as the one
developed by Socas-Navarro et al (1998; 2000a).
\nocite{SNRCTB98} 
\nocite{SNTBRC00a} 
Such detailed study,
however, is beyond the scope of the present work.

The observations reveal an intricate pattern of field/thermodynamic topologies, with
remarkable differences between the photosphere and the chromosphere. The
chromosphere itself looks different depending on whether we look at the red
or the blue magnetograms. The blue side of the Ca~II line cores is often
perturbed by the occurrence of strong dynamical phenomena, resulting in very
asymmetric ``anomalous'' polarization profiles (\citeNP{SNTBRC00b};
\citeNP{SNTBRC00c}; \citeNP{SNTBRC01}).

The red magnetograms from both Ca~II lines reveal a ``diffuse'' pattern
beyond the limb-side (right of the images) of both sunspots.
Interestingly, this pattern exhibits significant differences between
the two chrosmospheric lines. In some cases it even has the opposite
polarity. For instance near the top-right corner of the images 849.8 has
a mixture of black and white, whereas 854.2 is all white. 

The neutral line is clearly visible on the limb-side penumbra in the
photosphere (top-right panel in Fig~\ref{fig:maps}), exhibiting an
alternance of black and white filaments. However, the penumbral filaments in
the chromosphere all have the same polarity as the umbra (black). 

Notice what appear to be two network cells on the right edge (upper part of
the image) of the photospheric magnetogram. The 849.8 red magnetogram shows
the same structure as a chain of more isolated bright knots (some of them
with opposite Stokes~V polarity). These cells are
more difficult to discern in both (red and blue) 854.2 magnetograms. The
structure of the filaments extending out of the sunspot seems to merge, at
least visually, with these network cells. The upper cell interior, which is
empty in the photospheric magnetogram, is filled with a diffuse field in the
849.8 magnetogram. The 854.2 image shows isolated structures in the cell
interior. 

A small facula that appears as a sharp black feature at coordinates
($x=55$,$y=20$) in the photospheric magnetogram shows as a diffuse black area
in 849.8 surrounded by a white region. In 854.2 the black feature is even more
diffuse. At the very bottom of the photospheric image, near $x=60$, there is
a small positive (white) magnetic feature. In 849.8 this is clearly visible as
two bright knots, surrounded by opposite polarity field.

The area to the left of the sunspots is also interesting. The predominantly
negative polarity in the photosphere turns into a small-scale mixture of
small black and white dots. The 854.2 line, however, shows again a
concentration of predominantly negative polarity in the upper part.

Finally, notice how the separation between the two sunspots, which is about
5\arcsec \, in the photospheric magnetogram, becomes shorter in the 849.8
magnetogram and finally disappears when seen in 854.2.

The linear polarization signal, shown in Fig~\ref{fig:linmaps}, is only
prominent in the sunspots. The photospheric and chromospheric maps are rather
similar, although the chromospheric map is more diffuse and exhibits
filaments extending further away from the sunspot. 

A sample of the slit spectra recorded by SPINOR is presented in
Fig~\ref{fig:profs}. The data in the figure are from a scan of NOAA~0635
observed on June~19 at 14:08~UT. The problems with the Pluto camera computer
had been 
solved during the previous days, allowing us to record data at 1083~nm.
The signal-to-noise ratio 
is quite poor due to 1) the the low polarimetric efficiency discussed
above, and 2) especially to the low quantum efficiency of the camera at this
wavelength ($\simeq$3\%). The 1083~nm data displayed in the figure have been
binned in the 
scanning direction over a 1.5\arcsec \, interval. The 849.8 and 854.2~nm
data are unbinned (0.22\arcsec). 

\subsection{Velocities}

Photospheric velocity maps of NOAA~0634 were obtained by measuring the
minimum position of the Fe~I line at 849.7~nm. A similar strategy is not
viable for the Ca~II lines, however, because of the complicated
patterns of emission reversals and self-absortions found in the line
cores. Instead we measured the intensity difference between two points
symmetrically located at various distances from the line core. This method is
only an approximation but it seems to work better than taking the
minimum position. 

\begin{figure*}
\includegraphics[width=1.2\maxfloatwidth]{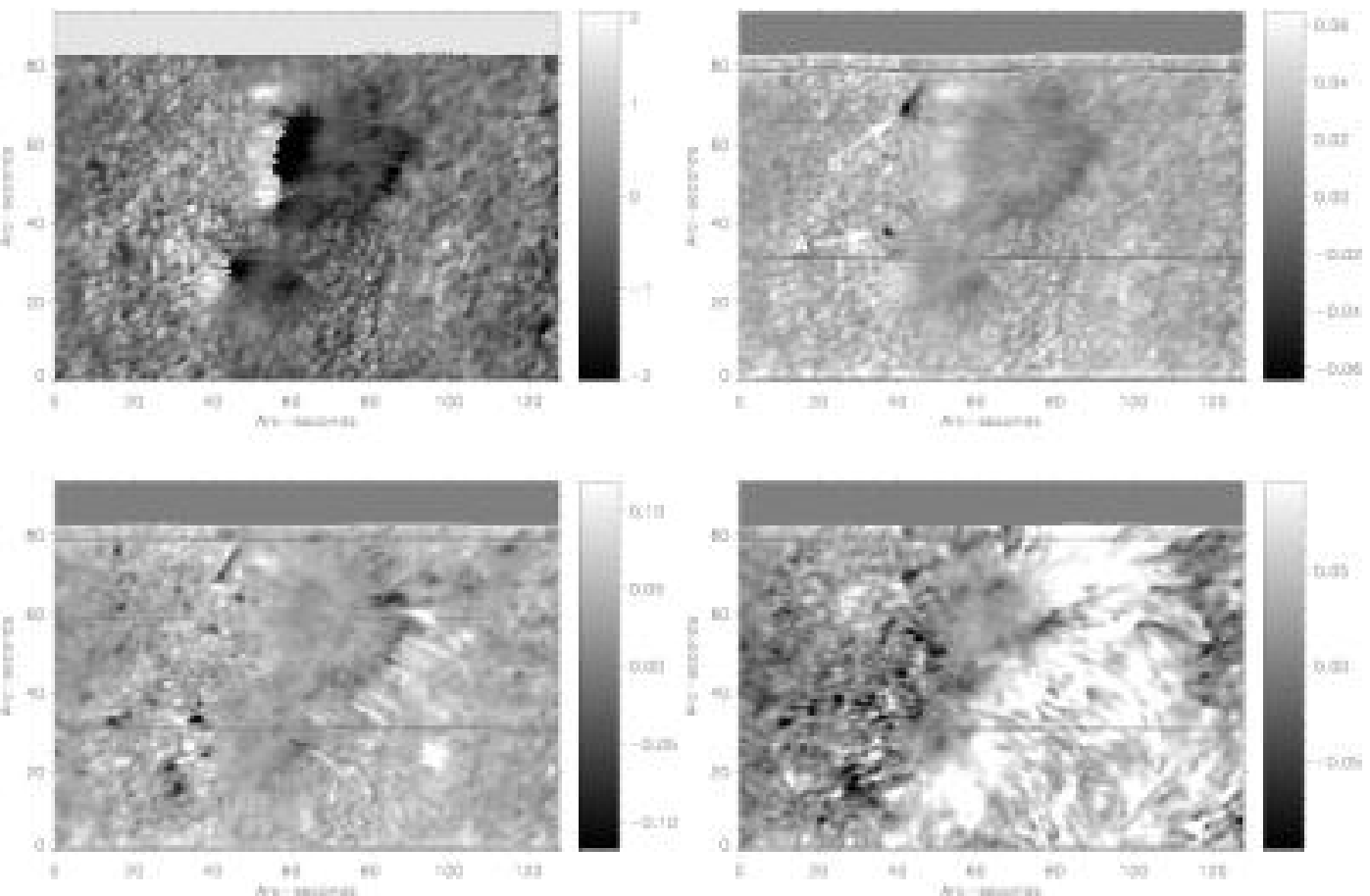}
\caption{
Upper left: Line of sight velocity measured in the core of a photospheric
Fe~I line 
(scale is km~s$^{-1}$). Other panels: Line of sight velocity maps measured as
the difference between blue and red intensity at 75, 37 and 7~pm from the
849.8~nm line center, in units of continuum intensity. In all cases
(black/white) represents flows (away from/towards) the observer. The arrows
mark the positions of two interesting features discussed in the text.(Note: Figure has been downsampled for Astro-ph)
}
\label{fig:velmaps}
\end{figure*}

\begin{figure*}
\includegraphics[width=1.2\maxfloatwidth]{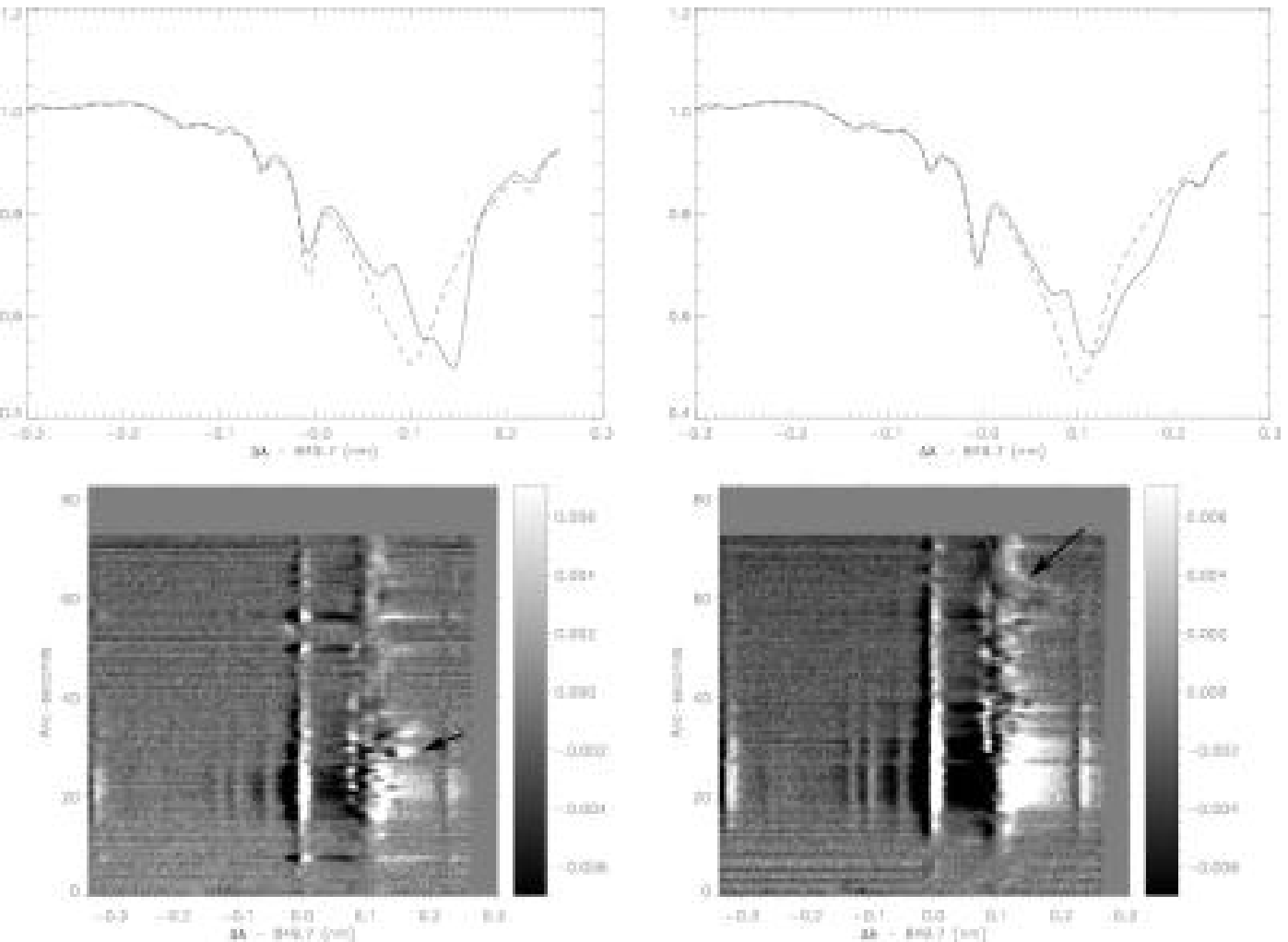}
\caption{
Upper panels: Stokes~$I$ profiles of points~A (left) and~B(right) normalized
to quiet Sun continuum. The solid line represents the average intensity
profile along the slit direction. Note the strong redshifts present in the
chromospheric lines. Lower panels: Stokes~$V$ spectra at the slit positions
of points~A (left) and~B (right) in Fig~5. The arrows mark the positions
along the slit of these two points. (Note: Figure has been downsampled for Astro-ph)
}
\label{fig:velprofs}
\end{figure*}

Fig~\ref{fig:velmaps} shows the velocity maps at various optical
heights. The cosine of the heliocentric angle for this region is
$\mu=0.74$. Notice the two prominent dark features marked with arrows in the 
figure. These features are very strong plasma flows directed away from the
line of 
sight and located on the center side of the sunspot (but further away than
the photospheric penumbra). Detailed profiles of these features are shown in
Fig~\ref{fig:velprofs}.

Both features exhibit very strong redshifts, visible in Stokes~$I$
and~$V$. Point~A shows a very prominent redshift starting near 40~pm away
from line center (corresponding roughly to the high photosphere or low
chromosphere) all the way to the line core (middle chromosphere). This
feature is well localized (see lower panel) and the Doppler shift is
$\sim$14~km~s$^{-1}$. Point~B exhibits somewhat weaker flows
($\sim$10~km~s$^{-1}$) flows that are more localized in
height (upper photosphere/lower chromosphere), but has a more extended tail
along the slit (see lower panel).

The polarization signal indicates that these features are of magnetic nature,
perhaps the result of chromospheric plasma falling down at high speed inside
a magnetic filament. Further observations with time resolution
and detailed Non-LTE modeling would be desirable to understand the structure
and evolution of such strong downflows.

\section{Conclusions}
\label{sec:conc}

This paper presents some initial results from first-light observations with
SPINOR, a new facility instrument for the DST.
We perceive an urgent demand in the solar community for a new
experiment-oriented  (as opposed 
to specialized) spectro-polarimeter. We expect that SPINOR will fulfill this
need and remain at the cutting-edge of solar research at least until the
construction of the Advanced Technology Solar Telescope
(\citeNP{KRH+02}). Its broad wavelength coverage will provide a 
uniquely connected view of photospheric, chromospheric and, to some extent,
coronal magnetism. The observations presented in the present paper are
intended to emphasize this point and to demonstrate the capabilities of the
instrument in the near infrared.

It is important to remark that the final configuration of the instrument will
provide significant improvements for infrared observations with respect to
the performance reported here. The new achromatic modulator (already
deployed) and cameras (larger format and more efficient) will hopefully
enhance SPINOR's performance at 1$\mu$m and extend its range up to 1.6
$\mu$m. 

As with the ASP, researchers from national and foreign institutions 
will be able to access the new instrument through the usual time allocation
competition for the Dunn Telescope, operated by NSO.

\section{Acknowledgments}

The authors wish to acknowledge the enthusiastic support from the NSO staff
at the Sac Peak observatory, especially D. Gilliam, M. Bradford and J.
Elrod. Thanks are also due to S. Hegwer, S. Gregory, R. Dunbar, T. Spence,
S. Fletcher, C. Berst and W. Jones.


\end{article}

\end{document}